\documentclass[11pt,epsf]{article}
\usepackage{graphics} 
\usepackage{amssymb} 
\usepackage{theorem}  
\usepackage{latexsym}  
 
\makeatletter
  \renewcommand{\section}{\@startsection
  {section}%
  {1}%
  {0mm}%
  {\baselineskip}%
  {0.5\baselineskip}%
  {\bfseries}}%
\makeatother

\makeatletter
  \renewcommand{\subsection}{\@startsection
  {subsection}%
  {2}%
  {0mm}%
  {\baselineskip}%
  {0.5\baselineskip}%
  {\bfseries \itshape}}%
\makeatother

\makeatletter
  \renewcommand{\subsubsection}{\@startsection
  {subsubsection}%
  {3}%
  {0mm}%
  {\baselineskip}%
  {0.5\baselineskip}%
  {\itshape}}%
\makeatother

\makeatletter
  \renewcommand{\@seccntformat}[1]{%
    \normalfont \bfseries \normalsize{\csname the#1\endcsname}.\hspace{0.5em}}
\makeatother

\newcommand{\vct}[1]{{\bf #1}}

\def\Max{\mathop{\mathrm{Max}}}
\def\Maxe{\mathop{\mathrm{Max_\epsilon}}}

\theorembodyfont{\rmfamily}
\newtheorem{definition}{Definition}[section]
\newtheorem{lemma}[definition]{\it Lemma}
\newtheorem{theorem}[definition]{\bfseries Theorem}

\theorembodyfont{\rmfamily}
\newtheorem{example}[definition]{Example}

\makeatletter
  
  \@addtoreset{definition}{section}
\makeatother

  

  

\makeatletter
  
  \@addtoreset{equation}{section}
\makeatother

\makeatletter 
  \renewcommand{\fnum@figure}[1]{%
  \footnotesize{Fig. \thefigure.}
  }
\makeatother

\begin{document}
\vspace*{0.8in}
\begin{center}
{\bfseries \Large
Inversible Max-Plus Algebras and \\
Integrable systems
}
\end{center}

\begin{center}
 {\footnotesize Tomoshiro OCHIAI}\\
 {\footnotesize \itshape Graduate School of 
 Mathematical Sciences, \\
 The University of Tokyo,\\
   Tokyo 153-8914, Japan}
\end{center}

\vspace{4cm}

\begin{abstract}
We present an extended version of max-plus algebra which includes the inverse operator of ``max". This algebra enables us to ultra-discretize the system including subtractions and obtain new ultra-discrete equations. The known ultra-discrete equations can also be recovered by this construction.
\end{abstract}

\begin{center}
PACS numbers: 02.30.Ik, 02.30.Jr, 05.45.Yv
\end{center}
 
\vspace{1in}

\newpage

\section{Introduction}
An integrable system is one of main subjects in mathematics and physics. Especially, soliton equations and integrable lattice models have been studied as integrable systems in  a variety of points of view. Recently, by the development of computer sciences, discrete and ultra-discrete versions of integral system have been attracting a great deal of attention. The ultra-discretization of soliton equations and other various important equations are intensively studied in \cite{Hirota, Hirota T, TakahashiSatsuma, Takahashi2, Tokihiro2, Tokihiro T M ,Tokihiro, Torii T S}. Moreover, it is known that the various ultra-discrete soliton equations are obtained from solvable vertex model in statistical mechanics in a certain limiting procedure \cite{Hatayama H I K T T, Hatayama K T}. The key formula for the ultra-discretization is given by 
\begin{eqnarray*}
\lim_{\epsilon \to \infty} \epsilon \log (e^{X/\epsilon}+e^{Y/\epsilon})
=\mbox{max}(X,Y)
\end{eqnarray*}
for arbitrary real numbers $X,Y$.
Using this formula, it is known that the field of real numbers can be transformed into the so-called "max-plus" algebra, which is well known in integrable systems. But the main difficulty for the usual ultra-discretization procedure is that there is no operator corresponding to the subtraction $``-"$. (see Fig.\ref{Fig:ultradiscretization}) Therefore, so far, we cannot deal with ultra-discretization of integral systems including the subtraction.

 In this paper, in order to solve this difficulty, we define an extended version of max-plus algebra called ``inversible Max-Plus algebra", which includes the usual max-plus algebra and the operator corresponding to the subtraction.

\begin{figure}[htbp]
	\begin{eqnarray*}
&&x \times y \longrightarrow X + Y \\
&&x / y \longrightarrow X - Y \\
&&x + y \longrightarrow \mbox{max}(X, Y) \\
&&x - y \longrightarrow ~~~~?? 
\end{eqnarray*}
	\caption{ultra-discretization}
	\label{Fig:ultradiscretization}
\end{figure}

Our constructions is as follows. We shall define a Max-Plus algebra as {\bf Z}-module $A$ with the operation $\mbox{Max}:A\times A\to A$ satisfying
\begin{eqnarray}
&&(\alpha+\beta)+\gamma=\alpha+(\beta+\gamma),\label{eqn: Max-Plus algebre 1}\\
&&\mbox{Max}(\mbox{Max}(\alpha,\beta),\gamma)=\mbox{Max}(\alpha,\mbox{Max}(\beta,\gamma)),\label{eqn: Max-Plus algebre 2}\\
&&\mbox{Max}(\alpha+\beta,\alpha+\gamma)=\alpha+\mbox{Max}(\beta,\gamma),
\label{eqn: Max-Plus algebre 3}
\end{eqnarray}
for $\alpha, \beta, \gamma \in A$.
Max-Plus subalgebra and homomorphism are defined in the usual manner. The basic example is ${\bf Z}$ which is the ring of integer with $\Max(n,m)=\max(n,m)$, where $\max(n,m)$ stands for the maximum of the pair $n$ and $m$.

In this paper, we explicitly construct Max-Plus algebra $\Omega$ and $\Omega_0$ with $P:\Omega_0\to {\bf Z}$ such that
\begin{list}{}{}
	\item (1) $\Omega\supset \Omega_0$ as Max-Plus algebra,
	\item (2) $P:\Omega_0\to {\bf Z}$ as Max-Plus homomorphism,
	\item (3) There exists an elment $\eta\in \Omega$ such that 
	\begin{eqnarray}\label{eqn; property}
	\mbox{Max}(\mbox{Max}(\alpha, \beta), \beta+\eta )=\alpha 
	\end{eqnarray}
	for arbitrary $\alpha, \beta\in \Omega$
\end{list}
We call $\Omega$ {\it inversible Max-Plus algebra}, due to the inversible property of Max operator (\ref{eqn; property}).

It is known that we can construct UD equation (ultra-discrete equation) on ${\bf Z}$ from standard discrete PDE's (partial differential equations) satisfying certain conditions through the usual ultra-discretization procedure. But, using our constructions, we can construct UD equation on $\Omega$ from any discrete PDE's, even if they include the subtraction. If this UD equation and its solution on $\Omega$ belong to $\Omega_0$, the projection $P$ gives the UD equation and solutions on ${\bf Z}$ explained above. As an application, we apply this inversible Max-Plus algebra $\Omega$ to periodic recursive equation and obtain UD periodic equation on $\Omega$ and ${\bf Z}$.
\\
 
\noindent{\bf Acknowledgements.} We would like to express our hearty thanks to Professor J.Satsuma for his advice and encouragement. 
 
\section{Inversible Max-Plus algebras} 
\subsection{Definition of $\tilde{\bf Z}$}
Let ${\bf Z}_2=\{0, \eta\}$ be the finite group of order two (i.e. $0+0=0$, $0+\eta=\eta$, $\eta+\eta=0$,) and set
\begin{eqnarray*}
\tilde{\bf Z}={\bf Z}\oplus {\bf Z}_2\cup \{-\infty\}=\{x+\xi ~|~ x\in {\bf Z}, \xi \in {\bf Z}_2\}\cup \{-\infty\}.
\end{eqnarray*}
Firstly, we shall define ``max" operator on $\tilde{\bf Z}$ (i.e., $\mbox{max}:\tilde{\bf Z}\times \tilde{\bf Z}\to \tilde{\bf Z}$) as follows. For $a=x+\xi \in {\bf Z}\oplus {\bf Z}_2$, $a^\prime=x^\prime+\xi^\prime \in {\bf Z}\oplus {\bf Z}_2$, set
\begin{eqnarray*}
&&\mbox{max}(a,-\infty)=a,\\
&&\mbox{max}(-\infty,-\infty)=-\infty,
\end{eqnarray*} 
and 
\begin{eqnarray*}
&&\mbox{max}(a,a^\prime)=\left\{
\begin{array}{ll}
a &(x > x^\prime),\\
a &(x = x^\prime~\mbox{and}~\xi = \xi^\prime),\\ 
a^\prime &(x<x^\prime),\\
-\infty &(x = x^\prime~\mbox{and}~\xi \neq  \xi^\prime).
\end{array}
\right.
\end{eqnarray*}

Secondly, we shall define ``plus" operator on $\tilde{\bf Z}$ (i.e., $\mbox{plus}:\tilde{\bf Z}\times \tilde{\bf Z}\to \tilde{\bf Z}$) as follows. For $a=x+\xi \in {\bf Z}\oplus {\bf Z}_2$ and $a^\prime=x^\prime+\xi^\prime \in {\bf Z}\oplus {\bf Z}_2$, set
\begin{eqnarray*}
&&\mbox{plus}(a,a^\prime)=(x+x^\prime)+(\xi+\xi^\prime),\\
&&\mbox{plus}(a,-\infty)=-\infty, \\
&&\mbox{plus}(-\infty,-\infty)=-\infty.   
\end{eqnarray*}
In the sequel, we write $a+b$ for $\mbox{plus}(a,b)$ for the matter of convenience.

If $a\neq a^\prime+\eta$, define the order relations $a>a^\prime$, $a<a^\prime$ and $a=a^\prime$ by
\begin{eqnarray*}
&&a>a^\prime \iff x > x^\prime,\\
&&a<a^\prime \iff x < x^\prime,\\
&&a=a^\prime \iff x = x^\prime~ \mbox{and} ~\xi =  \xi^\prime.
\end{eqnarray*}

If $a\neq \eta$, define the absolute value $|a|$ by
\begin{eqnarray*}
|a|=\mbox{max}(a,0)+\mbox{max}(-a,0).
\end{eqnarray*}

\subsection{Definition of ~$\Xi$}
Set
\begin{eqnarray*}
\tilde{\bf Z}^n=\{(a_1,\cdots,a_n)~|~a_1,\cdots,a_n \in \tilde{\bf Z}\}.
\end{eqnarray*}
For the matter of convenience, an element $(a_1,\cdots,a_n)$ is denoted by ${\bf a}$ or $(a_i)_{i=1}^{n}$.

Set $\Xi=\bigcup_{n=1}^\infty \tilde{\bf Z}^n/\sim$, where the union is disjoint and the equivalence relation "$\sim$" is generated by
\begin{eqnarray}
&&(\cdots,a_i,\cdots,a_j,\cdots)\sim (\cdots,a_j,\cdots,a_i,\cdots),
\label{eqn: equivalence relation1}\\
&&(a_1,\cdots,a_n,b,b+\eta)\sim (a_1,\cdots,a_n,-\infty),
\label{eqn: equivalence relation2}\\
&&(a_1,\cdots,a_n,-\infty)\sim (a_1,\cdots,a_n). \label{eqn: equivalence relation3}
\end{eqnarray}

An element $(a_1,\cdots,a_n)$ is called irreducible, if $n$ is minimum in its equivalence class. We remark that we can introduce the order relations into $a_1,\cdots,a_n$, if $(a_1,\cdots,a_n)$ is irreducible.

If there is no danger of confusion, the equivalence class defined by $(a_1,\cdots,a_n)$ is also denoted by $(a_1,\cdots,a_n)$.

Let us define "$\overline{\mbox{Max}}$" and "Plus" operators on $\Xi=\bigcup_{n=1}^\infty \tilde{\bf Z}^n/\sim$ (i.e., $\overline{\mbox{Max}}:\Xi \times \Xi \to \Xi$ and $\mbox{Plus}:\Xi \times \Xi \to \Xi$) as follows. For $\vct{a}=(a_i)_{i=1}^n=(a_1,\cdots,a_n)\in \Xi$ and $\vct{b}=(b_i)_{i=1}^m=(b_1,\cdots,b_m)\in \Xi$, set
\begin{eqnarray}
\overline{\mbox{Max}}(\vct{a},\vct{b})&=&(a_1,\cdots,a_n,b_1,\cdots,b_m),
\label{eqn:definition of bar Max}\\
\mbox{Plus}(\vct{a},\vct{b})&=&(a_i+b_j)_{i=1}^n{}_{j=1}^m\nonumber\\
&=&(a_1+b_j,\cdots,a_n+b_j)_{j=1}^m\nonumber\\
&=&(a_1+b_1,a_2+b_1,\cdots,a_n+b_1,a_1+b_2,\cdots,a_n+b_m). \label{eqn:definition of Plus}
\end{eqnarray}
In the sequel, we write $\vct{a}+\vct{b}$ for $\mbox{Plus}(\vct{a},\vct{b})$ for the matter of convenience.

\begin{lemma}
The above definition is well defined. 
\end{lemma}  
\noindent{\it Proof.}~~
It is enough to show that the definition is invariant under relations
(\ref{eqn: equivalence relation1}) $\sim$ (\ref{eqn: equivalence relation3}).

Firstly, we will show the "$\overline{\mbox{Max}}$" operator is invariant under relations (\ref{eqn: equivalence relation1}), (\ref{eqn: equivalence relation2}) and (\ref{eqn: equivalence relation3}). For $\vct{a}=(a_1\cdots,a_i,\cdots,a_j,\cdots,a_n)$, $\vct{a}^\prime=(a_1,\cdots,a_j,\cdots,a_i,\cdots,a_n)$ and $\vct{b}=(b_1,\cdots,b_m)$, we have
\begin{eqnarray*}
\overline{\mbox{Max}}(\vct{a},\vct{b})
&=&(a_1\cdots,a_i,\cdots,a_j,\cdots,a_n,b_1,\cdots,b_m)\\
&=&(a_1,\cdots,a_j,\cdots,a_i,\cdots,a_n,b_1,\cdots,b_m)\\
&=&\overline{\mbox{Max}}(\vct{a}^\prime,\vct{b}),
\end{eqnarray*}
which shows the invariance under relation (\ref{eqn: equivalence relation1}).
For $\vct{a}=(a_1\cdots,a_n,c,c+\eta)$, $\vct{a}^\prime=(a_1,\cdots,a_n,-\infty)$ and $\vct{b}=(b_1,\cdots,b_m)$, we have
\begin{eqnarray*}
\overline{\mbox{Max}}(\vct{a},\vct{b})
&=&(a_1\cdots,a_n,c,c+\eta,b_1,\cdots,b_m)\\
&=&(a_1,\cdots,a_n,-\infty,b_1,\cdots,b_m)\\
&=&\overline{\mbox{Max}}(\vct{a}^\prime,\vct{b}),
\end{eqnarray*}
which shows the invariance under relation (\ref{eqn: equivalence relation2}).
For $\vct{a}=(a_1\cdots,a_n,-\infty)$, $\vct{a}^\prime=(a_1,\cdots,a_n)$ and $\vct{b}=(b_1,\cdots,b_m)$, we have 
\begin{eqnarray*} 
\overline{\mbox{Max}}(\vct{a},\vct{b})&=&(a_1\cdots,a_n,-\infty,b_1,\cdots,b_m)\\
&=&(a_1,\cdots,a_n,b_1,\cdots,b_m)\\ 
&=&\overline{\mbox{Max}}(\vct{a}^\prime,\vct{b}),
\end{eqnarray*}
which shows the invariance under relation (\ref{eqn: equivalence relation3}).

Secondly, we will show the ``Plus" operator is invariant under relation (\ref{eqn: equivalence relation1}), (\ref{eqn: equivalence relation2}) and (\ref{eqn: equivalence relation3}). For $\vct{a}=(a_1\cdots,a_i,\cdots,a_j,\cdots,a_n)$, $\vct{a}^\prime=(a_1,\cdots,a_j,\cdots,a_i,\cdots,a_n)$ and $\vct{b}=(b_1,\cdots,b_m)$, we have
\begin{eqnarray*}
\vct{a}+\vct{b}&=&(a_1+b_k,\cdots,a_i+b_k,\cdots,a_j+b_k,\cdots,a_n+b_k)_{k=1}^m\\
&=&(a_1+b_k,\cdots,a_j+b_k,\cdots,a_i+b_k,\cdots,a_n+b_k)_{k=1}^m\\
&=&\vct{a}^\prime+\vct{b},
\end{eqnarray*}
which shows the invariance under relation (\ref{eqn: equivalence relation1}).
For $\vct{a}=(a_1\cdots,a_n,c,c+\eta)$, $\vct{a}^\prime=(a_1,\cdots,a_n,-\infty)$ and $\vct{b}=(b_1,\cdots,b_m)$, we have 
\begin{eqnarray*}
\vct{a}+\vct{b}&=&(a_1+b_k,\cdots,a_n+b_k,c+b_k,c+\eta+b_k)_{k=1}^m\\
&=&(a_1+b_k,\cdots,a_n+b_k,-\infty+b_k)_{k=1}^m\\
&=&\vct{a}^\prime+\vct{b},
\end{eqnarray*}
which shows the invariance under relation (\ref{eqn: equivalence relation2}).
For $\vct{a}=(a_1\cdots,a_n,-\infty)$, $\vct{a}^\prime=(a_1,\cdots,a_n)$ and $\vct{b}=(b_1,\cdots,b_m)$, we have 
\begin{eqnarray*}
\vct{a}+\vct{b}&=&(a_1+b_k,\cdots,a_n+b_k,-\infty+b_k)_{k=1}^m\\
&=&(a_1+b_k,\cdots,a_n+b_k)_{k=1}^m\\
&=&\vct{a}^\prime+\vct{b},
\end{eqnarray*}
which shows the invariance under relation (\ref{eqn: equivalence relation3}).
$\square$

\begin{lemma} For $\vct{a}, \vct{b}, \vct{c}\in \Xi$, we have the following identities:
\begin{eqnarray}
&&(\vct{a}+\vct{b})+\vct{c}=\vct{a}+(\vct{b}+\vct{c}),\label{eqn: formula 1}\\
&&\overline{\mbox{Max}}(\overline{\mbox{Max}}(\vct{a},\vct{b}),\vct{c})=\overline{\mbox{Max}}(\vct{a},\overline{\mbox{Max}}(\vct{b},\vct{c})),\label{eqn: formula 2}\\
&&\overline{\mbox{Max}}(\vct{a}+\vct{b},\vct{a}+\vct{c})=\vct{a}+\overline{\mbox{Max}}(\vct{b},\vct{c}),\label{eqn: formula 3}\\
&&\overline{\mbox{Max}}(\overline{\mbox{Max}}(\vct{a}, \vct{b}), \vct{b}+\eta )=\vct{a}. \label{eqn: formula 4}
\end{eqnarray}
\end{lemma}
\noindent{\it Proof.}~~ Suppose $\vct{a}=(a_1,\cdots,a_n)$ , $\vct{b}=(b_1,\cdots,b_m)$ and $\vct{c}=(c_1,\cdots,c_l)$. The proof is straightforward from (\ref{eqn:definition of bar Max}) and (\ref{eqn:definition of Plus}). First formula (\ref{eqn: formula 1}) is obtained by the following computation:
\begin{eqnarray*}
(\vct{a}+\vct{b})+\vct{c}=(a_i+b_j+c_k)_{i=1}^{n}{}_{j=1}^{m}{}_{k=1}^{l}=
\vct{a}+(\vct{b}+\vct{c}).
\end{eqnarray*}
Second formula (\ref{eqn: formula 2}) is obtained by the following computation:
\begin{eqnarray*}
\overline{\mbox{Max}}(\overline{\mbox{Max}}(\vct{a},\vct{b}),\vct{c})=(a_1\cdots,a_n,b_1,\cdots,b_m,c_1,\cdots,c_l)=
\overline{\mbox{Max}}(\vct{a},\overline{\mbox{Max}}(\vct{b},\vct{c})).
\end{eqnarray*}
Third formula (\ref{eqn: formula 3}) is obtained by the following computation:
\begin{eqnarray*}
\vct{a}+\overline{\mbox{Max}}(\vct{b},\vct{c})&=&\vct{a}+(b_1,\cdots,b_m,c_1,\cdots,c_l)\\
&=&(b_1+a_i,\cdots,b_m+a_i,c_1+a_i,\cdots,c_l+a_i)_{i=1}^{n}\\
&=&\overline{\mbox{Max}}((b_1+a_i,\cdots,b_m+a_i)_{i=1}^{n}, (c_1+a_i,\cdots,c_l+a_i)_{i=1}^{n})\\
&=&\overline{\mbox{Max}}(\vct{a}+\vct{b}, \vct{a}+\vct{c}).
\end{eqnarray*}
Last formula (\ref{eqn: formula 4}) is obtained by the following computation:
\begin{eqnarray*}
\overline{\mbox{Max}}(\overline{\mbox{Max}}(\vct{a}, \vct{b}), \vct{b}+\eta )&=&(a_1\cdots,a_n,b_1,\cdots,b_m,b_1+\eta,\cdots,b_m+\eta)\\
&=&(a_1\cdots,a_n)\\
&=&\vct{a}.
\end{eqnarray*}
$\square$

\subsection{Definition of $\Phi$}
 Set $\Phi=\Xi \times \Xi^\prime/\sim$, where $\Xi^\prime=\Xi\backslash\{(-\infty)\}$ and the equivalence relation is given by
\begin{eqnarray}\label{eqn: equivalence relation on Xi}
(\vct{a},\vct{b})\sim (\vct{a}^\prime, \vct{b}^\prime) \iff \vct{a}+\vct{b}^\prime=\vct{a}^\prime+\vct{b},
\end{eqnarray} 
for $(\vct{a},\vct{b}),(\vct{a}^\prime, \vct{b}^\prime)\in \Xi \times \Xi^\prime$.
The equivalence class of $(\vct{a},\vct{b})$ is denoted by $\vct{a}-\vct{b}$. Then, we have
\begin{eqnarray*}
\Phi=\{\vct{a}-\vct{b}~|~ \vct{a}\in \Xi, \vct{b}\in \Xi^\prime \}.
\end{eqnarray*}
Next, we will define "$\mbox{Max}$" , "Plus" and "Minus" operators on $\Phi$ (i.e., $\mbox{Max}:\Phi \times \Phi \to \Phi$, $\mbox{Plus}:\Phi \times \Phi \to \Phi$ and $\mbox{Minus}:\Phi \times \Phi \to \Phi$) as follows. 
 For $\alpha=\vct{a}-\vct{b}\in \Phi$ and $\alpha^\prime=\vct{a}^\prime-\vct{b}^\prime \in \Phi$, set 
\begin{eqnarray}
&&\mbox{Max}(\alpha,\alpha^\prime)=\overline{\mbox{Max}}(\vct{a}+\vct{b}^\prime, \vct{a}^\prime+\vct{b})-(\vct{b}+\vct{b}^\prime), \label{eqn: definition of extended max}\\
&&\mbox{Plus}(\alpha,\alpha^\prime)=(\vct{a}+\vct{a}^\prime)-(\vct{b}+\vct{b}^\prime),\\
&&\mbox{Minus}(\alpha,\alpha^\prime)=(\vct{a}+\vct{b}^\prime)-(\vct{b}+\vct{a}^\prime).
\end{eqnarray}
In the sequel, for the matter of convenience, we write $\alpha+\alpha^\prime$ and $\alpha-\alpha^\prime$ for $\mbox{Plus}(\alpha,\alpha^\prime)$ and $\mbox{Minus}(\alpha,\alpha^\prime)$ respectively.

\begin{lemma}~~
The above definition is well defined. 
\end{lemma}
\noindent{\it Proof.}~~Firstly, we will show the ``Max" operator is invariant under relations (\ref{eqn: equivalence relation on Xi}). For $\vct{a}-\vct{b}=\vct{a}^\prime-\vct{b}^\prime\in \Phi$ and $\vct{c}-\vct{d}\in \Phi$, we have
\begin{eqnarray*}
\mbox{Max}(\vct{a}-\vct{b},\vct{c}-\vct{d})&=&\overline{\mbox{Max}}(\vct{a}+\vct{d}, \vct{c}+\vct{b})-(\vct{b}+\vct{d})\\
&=&\overline{\mbox{Max}}(\vct{a}+\vct{b}^\prime+\vct{d}, \vct{c}+\vct{b}+\vct{b}^\prime)-(\vct{b}+\vct{b}^\prime+\vct{d})\\
&=&\overline{\mbox{Max}}(\vct{a}^\prime+\vct{b}+\vct{d}, \vct{c}+\vct{b}+\vct{b}^\prime)-(\vct{b}+\vct{b}^\prime+\vct{d})\\
&=&\overline{\mbox{Max}}(\vct{a}^\prime+\vct{d}, \vct{c}+\vct{b}^\prime)-(\vct{b}^\prime+\vct{d})\\
&=&\mbox{Max}(\vct{a}^\prime-\vct{b}^\prime,\vct{c}-\vct{d}),
\end{eqnarray*}
which shows the invariance under relation (\ref{eqn: equivalence relation on Xi}).

Secondly, we will show the ``Plus" operator is invariant under relations (\ref{eqn: equivalence relation on Xi}). For $\vct{a}-\vct{b}=\vct{a}^\prime-\vct{b}^\prime\in \Phi$ and $\vct{c}-\vct{d}\in \Phi$, we have
\begin{eqnarray*}
(\vct{a}-\vct{b})+(\vct{c}-\vct{d})&=&(\vct{a}+\vct{c})-(\vct{b}+\vct{d})\\
&=&(\vct{a}+\vct{b}^\prime+\vct{c})-(\vct{b}+\vct{b}^\prime+\vct{d})\\
&=&(\vct{a}^\prime+\vct{b}+\vct{c})-(\vct{b}+\vct{b}^\prime+\vct{d})\\
&=&(\vct{a}^\prime+\vct{c})-(\vct{b}^\prime+\vct{d})\\
&=&(\vct{a}^\prime-\vct{b}^\prime)+(\vct{c}-\vct{d}),
\end{eqnarray*}
which shows the invariance under relation (\ref{eqn: equivalence relation on Xi}).

Thirdly, we will show the ``Minus" operator is invariant under relations (\ref{eqn: equivalence relation on Xi}). For $\vct{a}-\vct{b}=\vct{a}^\prime-\vct{b}^\prime\in \Phi$ and $\vct{c}-\vct{d}\in \Phi$, we have
\begin{eqnarray*}
(\vct{a}-\vct{b})-(\vct{c}-\vct{d})&=&(\vct{a}+\vct{d})-(\vct{b}+\vct{c})\\
&=&(\vct{a}+\vct{b}^\prime+\vct{d})-(\vct{b}+\vct{b}^\prime+\vct{c})\\
&=&(\vct{a}^\prime+\vct{b}+\vct{d})-(\vct{b}+\vct{b}^\prime+\vct{c})\\
&=&(\vct{a}^\prime+\vct{d})-(\vct{b}^\prime+\vct{c})\\
&=&(\vct{a}^\prime-\vct{b}^\prime)-(\vct{c}-\vct{d}),
\end{eqnarray*}
which shows the invariance under relation (\ref{eqn: equivalence relation on Xi}).
$\square$

Next, we shall show that $\Phi$ has the Max-Plus algebra structures (\ref{eqn: Max-Plus algebre 1}), (\ref{eqn: Max-Plus algebre 2}), (\ref{eqn: Max-Plus algebre 3})  and the inversible property (\ref{eqn; property}). More precisely, we have the following lemma.

\begin{lemma} \label{lemma: Phi structure}
For $\alpha, \beta, \gamma \in \Phi$, we have the following identities: 
\begin{eqnarray}
&&(\alpha+\beta)+\gamma=\alpha+(\beta+\gamma),\label{eqn:identity in Phi 1}\\
&&\mbox{Max}(\mbox{Max}(\alpha,\beta),\gamma)=\mbox{Max}(\alpha,\mbox{Max}(\beta,\gamma)),\label{eqn:identity in Phi 2}\\
&&\mbox{Max}(\alpha+\beta,\alpha+\gamma)=\alpha+\mbox{Max}(\beta,\gamma),
\label{eqn:identity in Phi 3}\\
&&\mbox{Max}(\mbox{Max}(\alpha, \beta), \beta+\eta )=\alpha.\label{eqn:identity in Phi 4}
\end{eqnarray}
\end{lemma}
\noindent{\it Proof.}~~Suppose $\alpha=\vct{a}-\vct{b}$, $\beta=\vct{c}-\vct{d}$ and $\gamma=\vct{e}-\vct{f}$. First identity (\ref{eqn:identity in Phi 1}) is obtained by the following computation:
\begin{eqnarray*}
(\alpha+\beta)+\gamma&=&(\vct{a}+\vct{c}+\vct{e})-(\vct{b}+\vct{d}+\vct{f})\\
&=&\alpha+(\beta+\gamma).
\end{eqnarray*}
Second identity (\ref{eqn:identity in Phi 2}) is obtained by the following computation:
\begin{eqnarray*}
\mbox{Max}(\mbox{Max}(\alpha,\beta),\gamma)&=&
\mbox{Max}(\mbox{Max}(\vct{a}-\vct{b},\vct{c}-\vct{d}),\vct{e}-\vct{f})\\
&=&\mbox{Max}(\overline{\mbox{Max}}(\vct{a}+\vct{d},\vct{c}+\vct{b})-(\vct{b}+\vct{d}),\vct{e}-\vct{f})\\
&=&\overline{\mbox{Max}}(\overline{\mbox{Max}}(\vct{a}+\vct{d},\vct{c}+\vct{b})+\vct{f},\vct{e}+\vct{b}+\vct{d}))-(\vct{b}+\vct{d}+\vct{f})\\
&=&\overline{\mbox{Max}}(\overline{\mbox{Max}}(\vct{a}+\vct{d}+\vct{f}, \vct{c}+\vct{b}+\vct{f}), \vct{e}+\vct{b}+\vct{d})-(\vct{b}+\vct{d}+\vct{f})\\
&=&\overline{\mbox{Max}}(\vct{a}+\vct{d}+\vct{f}, \overline{\mbox{Max}}(\vct{c}+\vct{b}+\vct{f}, \vct{e}+\vct{b}+\vct{d}))-(\vct{b}+\vct{d}+\vct{f})\\
&=&\overline{\mbox{Max}}(\vct{a}+\vct{d}+\vct{f}, \overline{\mbox{Max}}(\vct{c}+\vct{f}, \vct{e}+\vct{d})+\vct{b})-(\vct{b}+\vct{d}+\vct{f})\\
&=&\mbox{Max}(\vct{a}-\vct{b}, \overline{\mbox{Max}}(\vct{c}+\vct{f}, \vct{e}+\vct{d})-(\vct{d}+\vct{f}))\\
&=&\mbox{Max}(\vct{a}-\vct{b}, \mbox{Max}(\vct{c}-\vct{d}, \vct{e}-\vct{f}))\\
&=&\mbox{Max}(\alpha,\mbox{Max}(\beta,\gamma)).
\end{eqnarray*}
Third identity (\ref{eqn:identity in Phi 3}) is obtained by the following computation:
\begin{eqnarray*}
\mbox{Max}(\alpha+\beta,\alpha+\gamma)&=&\mbox{Max}((\vct{a}-\vct{b})+(\vct{c}-\vct{d}),(\vct{a}-\vct{b})+(\vct{e}-\vct{f}))\\
&=&\mbox{Max}((\vct{a}+\vct{c})-(\vct{b}+\vct{d}),(\vct{a}+\vct{e})-(\vct{b}+\vct{f}))\\
&=&\overline{\mbox{Max}}((\vct{a}+\vct{c})+(\vct{b}+\vct{f}),(\vct{a}+\vct{e})+(\vct{b}+\vct{d}))-(\vct{b}+\vct{d}+\vct{b}+\vct{f})\\
&=&\vct{a}-\vct{b}+\overline{\mbox{Max}}(\vct{c}+\vct{f},\vct{e}+\vct{d}))-(\vct{d}+\vct{f})\\
&=&\vct{a}-\vct{b}+\mbox{Max}(\vct{c}-\vct{d},\vct{e}-\vct{f})\\
&=&\alpha+\mbox{Max}(\beta,\gamma).
\end{eqnarray*}
Final identity (\ref{eqn:identity in Phi 4}) is obtained by the following computation:
\begin{eqnarray*}
\mbox{Max}(\mbox{Max}(\alpha, \beta), \beta+\eta)&=&
\mbox{Max}(\mbox{Max}(\vct{a}-\vct{b}, \vct{c}-\vct{d}), \vct{c}-\vct{d}+\eta)\\
&=&\mbox{Max}(\overline{\mbox{Max}}(\vct{a}+\vct{d}, \vct{c}+\vct{b})-(\vct{b}+\vct{d}), \vct{c}+\eta-\vct{d})\\
&=&\overline{\mbox{Max}}(\overline{\mbox{Max}}(\vct{a}+\vct{d}, \vct{c}+\vct{b})+\vct{d}, \vct{c}+\eta+\vct{b}+\vct{d})-(\vct{b}+\vct{d}+\vct{d})\\
&=&\overline{\mbox{Max}}(\overline{\mbox{Max}}(\vct{a}+\vct{d}+\vct{d}, \vct{c}+\vct{b}+\vct{d}), \vct{c}+\vct{b}+\vct{d}+\eta)-(\vct{b}+\vct{d}+\vct{d})\\
&=&\vct{a}+\vct{d}+\vct{d}-(\vct{b}+\vct{d}+\vct{d})\\
&=&\vct{a}-\vct{b}\\
&=&\alpha.
\end{eqnarray*}
$\square$

\subsection{Definition of inversible Max-Plus algebra ~$\Omega$}
For the matter of convenience, a one-component element $\vct{a}=(a)\in \Phi$ is denoted by $a$, and $\vct{b}=(0,-b_1,\cdots,-b_n)\in \Phi$ is denoted by $[b_1,\cdots,b_n]$, if there is no danger of confusion.
\begin{lemma}\label{lemma; plus}
We have
\begin{eqnarray*}
[a_1,\cdots,a_n]+[b_1,\cdots,b_n]=[\{a_i\}_{i=1}^n, \{b_i\}_{i=1}^m, \{a_i+b_j\}_{i=1}^n{}_{j=1}^m].
\end{eqnarray*}
\end{lemma}
\noindent{\it Proof.} The computation is straightforward as follows:
\begin{eqnarray*} 
[a_1,\cdots,a_n]+[b_1,\cdots,b_n]
&=&(0,-a_1,\cdots,-a_n)+(0,-b_1,\cdots,-b_n)\\
&=&(0,\{-a_i\}_{i=1}^n, \{-b_i\}_{i=1}^m, \{-a_i-b_j\}_{i=1}^n{}_{j=1}^m)\\
&=&[\{a_i\}_{i=1}^n, \{b_i\}_{i=1}^m, \{a_i+b_j\}_{i=1}^n{}_{j=1}^m]. 
\end{eqnarray*}
$\square$

Define a subset $\Omega$ of $\Phi$ by
\begin{eqnarray}\label{eqn: Omega}
\Omega=\{a+[b_1,\cdots,b_n]-[c_1,\cdots,c_m]~|~ n, m \in {\bf N}, a\in \tilde{\bf Z}, b_i\in \tilde{\bf Z}_+, c_i\in \tilde{\bf Z}_+\},
\end{eqnarray}
where
\begin{eqnarray*}
\tilde{\bf Z}_+=\{x \in {\bf Z} | x \ge 0 \}\cup \{x+\eta \in  {\bf Z}\oplus {\bf Z}_2\ | x > 0 \}.
\end{eqnarray*}
For $a+[b_1,\cdots,b_n]-[c_1,\cdots,c_m] \in \Omega$, we call ``$a$" {\it a leading term} and ``$[b_1,\cdots,b_n]-[c_1,\cdots,c_m]$" {\it a correction term}.

\begin{lemma} We have
\begin{eqnarray*}
\Omega=\Phi.
\end{eqnarray*}
\end{lemma}
\noindent{\it Remark.}  $\Omega$ also has the Max-Plus algebra structures (\ref{eqn: Max-Plus algebre 1}), (\ref{eqn: Max-Plus algebre 2}), (\ref{eqn: Max-Plus algebre 3})  and the inversible property (\ref{eqn; property}), because of Lemma  \ref{lemma: Phi structure}.

\noindent{\it Proof.} By definition, it is clear that $\Omega \subset \Phi$. Therefore, it is enough to show that $\Phi\subset \Omega$. For $\alpha=(a_1,\cdots,a_n)-(b_1,\cdots,b_m)\in\Phi$, we may assume that $(a_1,\cdots,a_n)$ and $(b_1,\cdots,b_m)$ are irreducible and 
$a_1\ge \cdots \ge a_n$ and $b_1\ge \cdots\ge b_n$, without loss of generality. Then, we have
\begin{eqnarray*}
&&(a_1,\cdots,a_n)-(b_1,\cdots,b_m)\\
&=&a_1-b_1+(0,-(a_1-a_2),\cdots,-(a_1-a_n))-(0,-(b_1-b_2),\cdots,-(b_1-b_m))\\
&=&a_1-b_1+[(a_1-a_2),\cdots,(a_1-a_n)]-[(b_1-b_2),\cdots,(b_1-b_m)]
\in \Omega.
\end{eqnarray*}
Here, we have used $(a_1-a_i)\in \tilde{\bf Z}_+$ $(i=2,\cdots,n)$ and $(b_1-b_i)\in \tilde{\bf Z}_+$ $(i=2,\cdots,m)$.
$\square$

Next, in the following two theorems, we shall express Plus, Minus and Max operators in $\Omega$.
\begin{theorem}
For $\alpha=a+[b_1,\cdots,b_n]-[c_1,\cdots,c_m]\in\Omega$ and $\alpha^\prime=a^\prime+[b^\prime_1,\cdots,b^\prime_p]-[c^\prime_1,\cdots,c^\prime_q]\in\Omega$, the result of the computation of Plus and Minus operators is given by
\begin{eqnarray}
&&\alpha+\alpha^\prime=a+a^\prime+C_+(\alpha,\alpha^\prime),\label{eqn: main theorem3}\\
&&\alpha-\alpha^\prime=a-a^\prime+C_-(\alpha,\alpha^\prime),\label{eqn: main theorem4}
\end{eqnarray}
where the correction terms $C_+, C_-$ are given by
\begin{eqnarray*}
&&C_+(\alpha,\alpha^\prime)=[\{b_i\}_{i=1}^n, \{b_i^\prime\}_{i=1}^p, \{b_i+b_j^\prime\}_{i=1}^n{}_{j=1}^p]-[\{c_i\}_{i=1}^m, \{c_i^\prime\}_{i=1}^q, \{c_i+c_j^\prime\}_{i=1}^m{}_{j=1}^q],\\
&&C_-(\alpha,\alpha^\prime)=[\{b_i\}_{i=1}^n, \{c_i^\prime\}_{i=1}^q, \{b_i+c_j^\prime\}_{i=1}^n{}_{j=1}^q]-[\{c_i\}_{i=1}^m, \{b_i^\prime\}_{i=1}^p, \{c_i+b_j^\prime\}_{i=1}^m{}_{j=1}^p].
\end{eqnarray*}
\end{theorem}
\noindent{\it Proof.} The computation is straightforward from Lemma \ref{lemma; plus}. 
$\square$

Next, we express Max operator in $\Omega$. For arbitrary $\alpha=a+[b_1,\cdots,b_n]\in\Omega$ and $\alpha^\prime=a^\prime+[b^\prime_1,\cdots,b^\prime_m]\in\Omega$ with irreducible expressions (i.e. $[b_1,\cdots,b_n]$ and $[b^\prime_1,\cdots,b^\prime_m]$ are irreducible),  we may assume that $b_1\le \cdots \le b_n$ and $b^\prime_1 \le \cdots \le b^\prime_m$, without loss of generality. Then, we have the following theorem.
\begin{theorem}\label{theorem: main theorem}
(1) For $\alpha=a+[b_1,\cdots,b_n]\in\Omega$ and $\alpha^\prime=a^\prime+[b^\prime_1,\cdots,b^\prime_m]\in\Omega$ with $b_1\le \cdots \le b_n$ and $b^\prime_1 \le \cdots \le b^\prime_m$, the expression of $\mbox{Max}(\alpha,\alpha^\prime)$ is given as follows.

If $a \neq a^\prime+\eta$, we have
\begin{eqnarray}\label{eqn: main theorem1}
&&\mbox{Max}(\alpha,\alpha^\prime)=
\mbox{max}(a,a^\prime)+C_m(\alpha,\alpha^\prime),
\end{eqnarray}
where the correction term $C_m$ is given by
\begin{eqnarray*}
C_m(\alpha,\alpha^\prime)=\bigl[|a-a^\prime|, \{b_i+\mbox{max}(a^\prime-a,0)\}_{i=1}^{n}, \{b_i^\prime+\mbox{max}(a-a^\prime,0)\}_{i=1}^{m}\bigr].
\end{eqnarray*}

If $a = a^\prime+\eta$, we have
\begin{eqnarray}\label{eqn: main theorem2}
&&\mbox{Max}(\alpha,\alpha^\prime)=\left\{\begin{array}{l}
a-b_{k+1}+[\{b_{i}-b_{k+1}\}_{i=k+2}^{n}, \{b_{i}^\prime-b_{k+1}+\eta\}_{i=k+1}^{m}]\\
~~~~~~~~~~~\mbox{if}~~{}^\exists k < n,m ~~\mbox{s.t} ~~ b_i=b_i^{\prime}~~(1 \le i \le k) ~~\mbox{and}~~ b_{k+1}<b_{k+1}^\prime ,\\
a-b_{m+1}+[\{b_{i}-b_{m+1}\}_{i=m+2}^{n}]\\
~~~~~~~~~~~\mbox{if}~~b_i=b_i^{\prime}~~\mbox{for}~~1 \le i \le m < n,\\
h-\infty  \\
~~~~~~~~~~~\mbox{if}~~b_i=b_i^{\prime}~~\mbox{for}~~1 \le i \le m = n.\\
\end{array}\right.
\end{eqnarray}
\end{theorem}

\noindent{\it Remark 1.} 
We will extend the above formula for general cases as follows. For
\begin{eqnarray*}
&&\alpha=a+[b_1,\cdots,b_n]-[c_1,\cdots,c_m]\in\Omega,\\
&&\alpha^\prime=a^\prime+[b^\prime_1,\cdots,b^\prime_p]-[c^\prime_1,\cdots,c^\prime_q]\in\Omega,
\end{eqnarray*}
the expression of $\mbox{Max}(\alpha,\alpha^\prime)$ can be computed by using (\ref{eqn: main theorem1}), (\ref{eqn: main theorem2}) and the identity
\begin{eqnarray}\label{eqn: extended identity}
\mbox{Max}(\alpha,\alpha^\prime)=\mbox{Max}(a+[b]+[c^\prime], a^\prime+[b^\prime]+[c])-([c]+[c^\prime]).
\end{eqnarray}
Here, notice that the first term in the right hand side (\ref{eqn: extended identity}) can be computed by (\ref{eqn: main theorem1}), (\ref{eqn: main theorem2}). 

\noindent{\it Remark 2.} 
In (\ref{eqn: main theorem2}), for $[b_1,\cdots,b_n]$ and $[b^\prime_1,\cdots,b^\prime_m]$, we can assume that there are only the following three cases:
\begin{enumerate}
	\item There exists $k\in {\bf N}$ and $k < n,m$ such that $b_i=b_i^{\prime}$ $(i=1,\cdots,k)$ and $b_{k+1}<b_{k+1}^\prime$,
	\item $m < n$ and $b_i=b_i^{\prime}$~$(i=1,\cdots,m)$,
	\item $m = n$ and $b_i=b_i^{\prime}$~$(i=1,\cdots,m)$,
\end{enumerate}
without loss of generality. These three cases 1, 2 and 3 correspond to the first, second and third cases in (\ref{eqn: main theorem2}) respectively.

\noindent{\it Proof of Theorem \ref{theorem: main theorem}.}  
Let us consider the case $a \neq a^\prime+\eta$. Without loss of generality, we can assume that $a \ge a^\prime$. Then, we have
\begin{eqnarray*}
\mbox{Max}(\alpha,\alpha^\prime)&=&\mbox{Max}((a,a-b_1,\cdots,a-b_n),(a^\prime,a^\prime-b_1^\prime,\cdots,a^\prime-b_m^\prime))\\
&=&(a,a-b_1,\cdots,a-b_n,a^\prime,a^\prime-b_1^\prime,\cdots,a^\prime-b_m^\prime)\\
&=&a+(0,-b_1,\cdots,-b_n,-(a-a^\prime),-(b_1^\prime+a-a^\prime),\cdots,-(b_m^\prime+a-a^\prime))\\
&=&a+[a-a^\prime,b_1,\cdots,b_n,b_1^\prime+a-a^\prime,\cdots,b_m^\prime+a-a^\prime],
\end{eqnarray*}
which shows that (\ref{eqn: main theorem1}) holds for $a \ge a^\prime$. We can prove (\ref{eqn: main theorem1}) for the other case $a < a^\prime$ in the same way.

Next, let us consider the case $a = a^\prime+\eta$. Firstly, we consider the first case in (\ref{eqn: main theorem2}). (i.e. There exists $k\in {\bf N}$ and $k < n,m$ such that $b_i=b_i^{\prime}$ $(i=1,\cdots,k)$ and $b_{k+1}<b_{k+1}^\prime$.) In this case, we have
\begin{eqnarray*}
&&\mbox{Max}(\alpha,\alpha^\prime)\\
&=&(a,a-b_1,\cdots,a-b_n,a^\prime,a^\prime-b_1^\prime,\cdots,a^\prime-b_m^\prime)\\
&=&(a-b_{k+1},\cdots,a-b_n,a^\prime-b_{k+1}^\prime,\cdots,a^\prime-b_m^\prime)\\&=&a-b_{k+1}+(0,-(b_{k+2}-b_{k+1}),\cdots,-(b_n-b_{k+1}),-(b_{k+1}^\prime-b_{k+1}+\eta),\cdots,-(b_m^\prime-b_{k+1}+\eta))\nonumber\\
&=&a-b_{k+1}+[\{b_{i}-b_{k+1}\}_{i=k+2}^{n}, \{b_{i}^\prime-b_{k+1}+\eta\}_{i=k+1}^{m}],
\end{eqnarray*}
which shows the first case in (\ref{eqn: main theorem2}). 
Secondly, we consider the second case in (\ref{eqn: main theorem2}). (i.e. $m < n$ and $b_i=b_i^{\prime}$~$(i=1,\cdots,m)$.) In this case, we have
\begin{eqnarray*}
\mbox{Max}(\alpha,\alpha^\prime)
&=&(a,a-b_1,\cdots,a-b_n,a^\prime,a^\prime-b_1^\prime,\cdots,a^\prime-b_m^\prime)\\
&=&(a-b_{m+1},\cdots,a-b_n)\\
&=&a-b_{m+1}+(0,-(b_{m+2}-b_{m+1}),\cdots,-(b_n-b_{m+1}))\\
&=&a-b_{m+1}+[\{b_{i}-b_{m+1}\}_{i=m+2}^{n}],
\end{eqnarray*}
which shows the second case in (\ref{eqn: main theorem2}). 
Thirdly, we consider the third case in (\ref{eqn: main theorem2}). (i.e. $m = n$ and $b_i=b_i^{\prime}$~$(i=1,\cdots,m)$.) In this case, we have
\begin{eqnarray*}
\mbox{Max}(\alpha,\alpha^\prime)
&=&(a,a-b_1,\cdots,a-b_n,a^\prime,a^\prime-b_1^\prime,\cdots,a^\prime-b_n^\prime)\\
&=&-\infty,
\end{eqnarray*}
which shows the third case in (\ref{eqn: main theorem2}).  
$\square$

\section{Applications}
\subsection{The projection to the usual max-plus algebra {\bf Z}}
Let $\Omega_0$ be a subset of $\Omega$ defined by $\Omega_0=\{a+[b_1,\cdots,b_n]-[c_1,\cdots.c_m] \in \Omega ~|~a \in {\bf Z}\} \subset \Omega$. For $\alpha=a+[b_1,\cdots,b_n]-[c_1,\cdots.c_m] \in \Omega_0$, define a projection map $\mbox{P}:\Omega_0 \to \bf{Z}$ by 
\begin{eqnarray}
\mbox{P}(\alpha)=a.
\end{eqnarray}

\begin{lemma}
~~$\Omega_0$ is Max-Plus subalgebra.
\end{lemma}
\noindent{\it Proof.}
It is clear from (\ref{eqn: main theorem1}), (\ref{eqn: main theorem3}) and (\ref{eqn: main theorem4}). Notice that $a, a^\prime \in \bf{Z}$.
 
\begin{theorem}
The map $\mbox{P}:\Omega_0 \to \bf{Z}$ is a homomorphism map with respect to ``Max", ``Plus", ``Minus" operators. More precisely, for $\alpha=a+[b_1,\cdots,b_n]-[c_1,\cdots.c_m] \in \Omega_0 $ and $\alpha^\prime=a^\prime+[b^\prime_1,\cdots,b^\prime_p]-[c^\prime_1,\cdots,c^\prime_q] \in \Omega_0$, we have
\begin{eqnarray*}
\mbox{P}(\mbox{Max}(\alpha,\alpha^\prime))=\mbox{max}(a,a^\prime),\\
\mbox{P}(\alpha+\alpha^\prime)=a+a^\prime,\\
\mbox{P}(\alpha-\alpha^\prime)=a-a^\prime.
\end{eqnarray*}
\end{theorem} 
\noindent{\it Proof.}
It is clear from (\ref{eqn: main theorem1}), (\ref{eqn: main theorem3}) and (\ref{eqn: main theorem4}). Notice that $a, a^\prime \in \bf{Z}$.

\subsection{The faithful representation on $\Omega$}
For any positive real number $\epsilon$ and $a=x+\xi\in \tilde{\bf Z}$, set $\overline{a}=x+\overline{\xi}\in {\bf C}$, where 
\begin{eqnarray*}
\overline{\xi}=\left\{\begin{array}{ll}
i\pi\epsilon & (\xi=\eta), \\
0 & (\xi=0).
\end{array}
\right.
\end{eqnarray*}

For $\alpha=(a_i)_{i=1}^n-(b_i)_{i=1}^m=(a_1,\cdots,a_n)-(b_1,\cdots,b_m)\in \Phi$, define a map $\mbox{R}_\epsilon:\Phi \to {\bf C}$ by
\begin{eqnarray*}
\mbox{R}_\epsilon(\alpha)&=&\epsilon\log(\sum_{i}e^{\overline{a_i}/\epsilon})-\epsilon\log(\sum_{i}e^{\overline{b_i}/\epsilon})\\
&=&\epsilon\log(e^{\overline{a_1}/\epsilon}+\cdots+e^{\overline{a_n}/\epsilon})
-\epsilon\log(e^{\overline{b_1}/\epsilon}+\cdots+e^{\overline{b_m}/\epsilon}).
\end{eqnarray*}
Here we set $e^{\overline{a_i}/\epsilon}$ to be zero, if $a_i=-\infty$.

\begin{lemma}
The above definition $\mbox{R}_\epsilon:\Phi \to {\bf C}$ is well defined.
\end{lemma}
\noindent{\it Proof.} Firstly, we shall show that the definition is invariant under relations (\ref{eqn: equivalence relation1}), (\ref{eqn: equivalence relation2}) and (\ref{eqn: equivalence relation3}). For $\vct{a}=(a_1,\cdots,a_i,\cdots,a_j,\cdots,a_n)$, $\vct{a}^\prime=(a_1,\cdots,a_j,\cdots,a_i,\cdots,a_n)$ and $\vct{b}=(b_1,\cdots,b_m)$, we have
\begin{eqnarray*}
\mbox{R}_\epsilon(\vct{a}-\vct{b})
&=&\mbox{R}_\epsilon((a_1,\cdots,a_i,\cdots,a_j,\cdots,a_n)-(b_1,\cdots,b_m))\\
&=&\epsilon\log(e^{\overline{a_1}/\epsilon}+\cdots+e^{\overline{a_i}/\epsilon}+\cdots+e^{\overline{a_j}/\epsilon}+\cdots+e^{\overline{a_n}/\epsilon})
-\epsilon\log(e^{\overline{b_1}/\epsilon}+\cdots+e^{\overline{b_m}/\epsilon})\\
&=&\epsilon\log(e^{\overline{a_1}/\epsilon}+\cdots+e^{\overline{a_j}/\epsilon}+\cdots+e^{\overline{a_i}/\epsilon}+\cdots+e^{\overline{a_n}/\epsilon})
-\epsilon\log(e^{\overline{b_1}/\epsilon}+\cdots+e^{\overline{b_m}/\epsilon})\\
&=&\mbox{R}_\epsilon((a_1,\cdots,a_j,\cdots,a_i,\cdots,a_n)-(b_1,\cdots,b_m))\\
&=&\mbox{R}_\epsilon(\vct{a}^\prime-\vct{b}),
\end{eqnarray*}
which shows the invariance under relation (\ref{eqn: equivalence relation1}).
For $\vct{a}=(a_1\cdots,a_n,c,c+\eta)$, $\vct{a}^\prime=(a_1,\cdots,a_n,-\infty)$ and $\vct{b}=(b_1,\cdots,b_m)$, we have
\begin{eqnarray*}
\mbox{R}_\epsilon(\vct{a}-\vct{b})
&=&\mbox{R}_\epsilon((a_1,\cdots,a_n,c,c+\eta)-(b_1,\cdots,b_m))\\
&=&\epsilon\log(e^{\overline{a_1}/\epsilon}+\cdots+e^{\overline{a_n}/\epsilon}+e^{\overline{c}/\epsilon}+e^{\overline{c+\eta}/\epsilon})
-\epsilon\log(e^{\overline{b_1}/\epsilon}+\cdots+e^{\overline{b_m}/\epsilon})\\
&=&\epsilon\log(e^{\overline{a_1}/\epsilon}+\cdots+e^{\overline{a_n}/\epsilon}+e^{\overline{c}/\epsilon}+e^{\overline{c}/\epsilon+i\pi})
-\epsilon\log(e^{\overline{b_1}/\epsilon}+\cdots+e^{\overline{b_m}/\epsilon})\\
&=&\epsilon\log(e^{\overline{a_1}/\epsilon}+\cdots+e^{\overline{a_n}/\epsilon}+e^{\overline{c}/\epsilon}-e^{\overline{c}/\epsilon})
-\epsilon\log(e^{\overline{b_1}/\epsilon}+\cdots+e^{\overline{b_m}/\epsilon})\\
&=&\epsilon\log(e^{\overline{a_1}/\epsilon}+\cdots+e^{\overline{a_n}/\epsilon})
-\epsilon\log(e^{\overline{b_1}/\epsilon}+\cdots+e^{\overline{b_m}/\epsilon})\\
&=&\mbox{R}_\epsilon(\vct{a}^\prime-\vct{b}),
\end{eqnarray*} 
which shows the invariance under relation (\ref{eqn: equivalence relation2}).
For $\vct{a}=(a_1\cdots,a_n,-\infty)$, $\vct{a}^\prime=(a_1,\cdots,a_n)$ and $\vct{b}=(b_1,\cdots,b_m)$, we have 
\begin{eqnarray*}
\mbox{R}_\epsilon(\vct{a}-\vct{b})
&=&\mbox{R}_\epsilon((a_1,\cdots,a_n,-\infty)-(b_1,\cdots,b_m))\\
&=&\epsilon\log(e^{\overline{a_1}/\epsilon}+\cdots+e^{\overline{a_n}/\epsilon})
-\epsilon\log(e^{\overline{b_1}/\epsilon}+\cdots+e^{\overline{b_m}/\epsilon})\\
&=&\mbox{R}_\epsilon(\vct{a}^\prime-\vct{b}),
\end{eqnarray*} 
which shows the invariance under relation (\ref{eqn: equivalence relation3}).

Secondly, we shall show that the definition is invariant under relation (\ref{eqn: equivalence relation on Xi}). Suppose $(a_1,\cdots,a_n)-(b_1,\cdots,b_m)=(a_1^\prime,\cdots,a_p^\prime)-(b_1^\prime,\cdots,b_q^\prime)$. Then we have
\begin{eqnarray*}
(a_i+b_j^\prime)_{i=1}^n{}_{j=1}^q=(a_i^\prime+b_j)_{i=1}^p{}_{j=1}^m.
\end{eqnarray*}
Then, it follows that
\begin{eqnarray}
\epsilon\log(\sum_{i=1}^n \sum_{j=1}^q e^{\overline{a_i+b_j^\prime}/\epsilon})
=\epsilon\log(\sum_{i=1}^p \sum_{j=1}^m e^{\overline{a_i^\prime+b_j}/\epsilon}).\label{eqn: identity of R}
\end{eqnarray}
Then, we have
\begin{eqnarray*}
&&\mbox{R}_\epsilon((a_1,\cdots,a_n)-(b_1,\cdots,b_m))\\
&=&\epsilon\log(\sum_{i=1}^n e^{\overline{a_i}/\epsilon})
-\epsilon\log(\sum_{j=1}^m e^{\overline{b_j}/\epsilon})\\
&=&\epsilon\log(\sum_{i=1}^p e^{\overline{a_i^\prime}/\epsilon})
-\epsilon\log(\sum_{j=1}^q e^{\overline{b_j^\prime}/\epsilon})
+\epsilon\log(\sum_{i=1}^n \sum_{j=1}^q e^{\overline{a_i+b_j^\prime}/\epsilon})
-\epsilon\log(\sum_{i=1}^p \sum_{j=1}^m e^{\overline{a_i^\prime+b_j}/\epsilon})
\\
&=&\epsilon\log(\sum_{i=1}^p e^{\overline{a_i^\prime}/\epsilon})
-\epsilon\log(\sum_{j=1}^q e^{\overline{b_j^\prime}/\epsilon})\\
&=&\mbox{R}_\epsilon((a_1^\prime,\cdots,a_p^\prime)-(b_1^\prime,\cdots,b_q^\prime)),
\end{eqnarray*}
which shows the invariance under relation (\ref{eqn: equivalence relation on Xi}). Here, we have used (\ref{eqn: identity of R}) in the third line.
$\square$

For any positive real number $\epsilon$, define a map $\mbox{Max}_\epsilon:{\bf C} \times {\bf C} \to {\bf C}$ by
\begin{eqnarray*}
\mbox{Max}_\epsilon(a,b)=\epsilon\log(e^{a/\epsilon}+e^{b/\epsilon}),
\end{eqnarray*}
where $a, b\in {\bf C}$. Here, we remark that ${\bf C}$ with $\mbox{Max}_\epsilon$ has the Max-Plus algebra structures (\ref{eqn: Max-Plus algebre 1}), (\ref{eqn: Max-Plus algebre 2}) and (\ref{eqn: Max-Plus algebre 3}).

\begin{theorem}
(1) The map $\mbox{R}_\epsilon:\Phi \to {\bf C}$ is a representation of $\Phi$. More precisely, for $\alpha, \alpha^\prime \in \Phi$, we have
\begin{eqnarray}
\mbox{R}_\epsilon(\mbox{Max}(\alpha,\alpha^\prime))=\mbox{Max}_\epsilon(\mbox{R}_\epsilon(\alpha),\mbox{R}_\epsilon(\alpha^\prime)),\label{eqn:rep1}\\
\mbox{R}_\epsilon(\alpha+\alpha^\prime)=\mbox{R}_\epsilon(\alpha)+\mbox{R}_\epsilon(\alpha^\prime),\label{eqn:rep2}\\
\mbox{R}_\epsilon(\alpha-\alpha^\prime)=\mbox{R}_\epsilon(\alpha)-\mbox{R}_\epsilon(\alpha^\prime).\label{eqn:rep3}
\end{eqnarray}

(2) If $\mbox{R}_\epsilon(\alpha)=0$ for any $\epsilon$, then $\alpha=0$.
\end{theorem}

\noindent{\it Proof.} (1) Firstly, we will prove (\ref{eqn:rep1}) as follows. For $\alpha=\vct{a}-\vct{b}=(a_1,\cdots,a_n)-(b_1,\cdots,b_m)\in \Phi$  and $\alpha=\vct{a}^\prime-\vct{b}^\prime=(a_1^\prime,\cdots,a_p^\prime)-(b_1^\prime,\cdots,b_q^\prime) \in \Phi$, we have
\begin{eqnarray*}
\mbox{R}_\epsilon(\mbox{Max}(\alpha,\alpha^\prime))
&=&\mbox{R}_\epsilon(\overline{\mbox{Max}}(\vct{a}+\vct{b}^\prime, \vct{a}^\prime+\vct{b})-(\vct{b}+\vct{b}^\prime))\\
&=&\mbox{R}_\epsilon\Bigl(\bigl((a_i+b^\prime_j)_{i=1}^n{}_{j=1}^q, (a^\prime_k+b_l)_{k=1}^p{}_{l=1}^m\bigr)-(b_i+b^\prime_j)_{i=1}^m{}_{j=1}^q\Bigr)\\
&=&\epsilon\log(\sum_{i,j}e^{\overline{a_i+b_j^\prime}/\epsilon}+\sum_{i,j}e^{\overline{a_i^\prime+b_j}/\epsilon})-\epsilon\log(\sum_{i,j}e^{\overline{b_i+b_j^\prime}/\epsilon})\\
&=&\epsilon\log(\frac{\sum_{i}e^{\overline{a_i}/\epsilon}}{\sum_{i}e^{\overline{b_i}/\epsilon}}+\frac{\sum_{i}e^{\overline{a_i^\prime}/\epsilon}}{\sum_{i}e^{\overline{b_i^\prime}/\epsilon}})\\
&=&\mbox{Max}_\epsilon(\epsilon\log(\frac{\sum_{i}e^{\overline{a_i}/\epsilon}}{\sum_{i}e^{\overline{b_i}/\epsilon}}),\epsilon\log(\frac{\sum_{i}e^{\overline{a_i^\prime}/\epsilon}}{\sum_{i}e^{\overline{b_i^\prime}/\epsilon}}))\\
&=&\mbox{Max}_\epsilon(\epsilon\log(\sum_{i}e^{\overline{a_i}/\epsilon})-\epsilon\log({\sum_{i}e^{\overline{b_i}/\epsilon}}),\epsilon\log(\sum_{i}e^{\overline{a_i^\prime}/\epsilon})-\epsilon\log(\sum_{i}e^{\overline{b_i^\prime}/\epsilon})\\&=&\mbox{Max}_\epsilon(\mbox{R}_\epsilon(\alpha),\mbox{R}_\epsilon(\alpha^\prime)),
\end{eqnarray*}
which shows (\ref{eqn:rep1}). Secondly, we will prove (\ref{eqn:rep2}) as follows:
\begin{eqnarray*}
\mbox{R}_\epsilon(\mbox{Max}(\alpha+\alpha^\prime))
&=&\mbox{R}_\epsilon((\vct{a}+\vct{a}^\prime)-(\vct{b}+\vct{b}^\prime))\\
&=&\mbox{R}_\epsilon\Bigl(\bigl((a_i+a^\prime_j)_{i=1}^n{}_{j=1}^p-(b_i+b^\prime_j)_{i=1}^m{}_{j=1}^q\Bigr)\\
&=&\epsilon\log(\sum_{i,j}e^{\overline{a_i+a_j^\prime}/\epsilon})-\epsilon\log(\sum_{i,j}e^{\overline{b_i+b_j^\prime}/\epsilon})\\
&=&\epsilon\log(\sum_{i}e^{\overline{a_i}/\epsilon})-\epsilon\log(\sum_{i}e^{\overline{b_i}/\epsilon})+\epsilon\log(\sum_{i}e^{\overline{a_i^\prime}/\epsilon})-\epsilon\log(\sum_{i}e^{\overline{b_i^\prime}/\epsilon})\\
&=&\mbox{R}_\epsilon\Bigl(\bigl((a_i)_{i=1}^n-(b_i)_{i=1}^m\Bigr)+\mbox{R}_\epsilon\Bigl(\bigl((a^\prime_i)_{i=1}^n-(b^\prime_j)_{i=1}^q\Bigr)\\
&=&\mbox{R}_\epsilon(\alpha)+\mbox{R}_\epsilon(\alpha^\prime).
\end{eqnarray*}
This shows (\ref{eqn:rep2}). We can obtain (\ref{eqn:rep3}) in the same way.

(2) For $\alpha=(a_1,\cdots,a_n)-(b_1,\cdots,b_m)\in\Phi$, it is assumed that $\alpha$ is irreducible and 
$a_1\ge\cdots\ge a_n$ and $b_1\ge\cdots\ge b_n$. Then, we have
\begin{eqnarray*}
\mbox{R}_\epsilon(\alpha)
&=&\epsilon\log(e^{\overline{a_1}/\epsilon}+\cdots+e^{\overline{a_n}/\epsilon})
-\epsilon\log(e^{\overline{b_1}/\epsilon}+\cdots+e^{\overline{b_m}/\epsilon})\\
&=&0.
\end{eqnarray*}
Hence, we have
\begin{eqnarray*}
e^{\overline{a_1}/\epsilon}+\cdots+e^{\overline{a_n}/\epsilon}=e^{\overline{b_1}/\epsilon}+\cdots+e^{\overline{b_m}/\epsilon}.
\end{eqnarray*}
This equation is equal to
\begin{eqnarray*}
e^{\overline{a_1}/\epsilon}(1+e^{-\overline{(a_1-a_2)}/\epsilon}+\cdots+e^{-\overline{(a_1-a_n)}/\epsilon})=e^{\overline{b_1}/\epsilon}(1+e^{-\overline{(b_1-b_2)}/\epsilon}+\cdots+e^{-\overline{(b_1-b_m)}/\epsilon}).
\end{eqnarray*}
Considering $\epsilon \to  \infty$, we have $a_1=b_1$. By using inductive argument, we obtain 
\begin{eqnarray*}
(a_1,\cdots,a_n)=(b_1,\cdots,b_m),
\end{eqnarray*}
which shows $\alpha=0$.
$\square$

\subsection{The ultra-discrete recursive equation}   
Define $\mbox{UD}:{\bf Z}\to \Omega$ by
\begin{eqnarray*}
\mbox{UD}(n)=\left\{\begin{array}{ll}
0+[\underbrace{0,\cdots,0}_{k-1}], &\mbox{if}~~ n=k, \\
-\infty &\mbox{if}~~ n=0, \\
\eta+[\underbrace{0,\cdots,0}_{k-1}] &\mbox{if}~~ n=-k. \\
\end{array}
\right.
\end{eqnarray*}

Let $P:{\bf R}^n \to {\bf R}$ be a polynomial function of finite degree given by
\begin{eqnarray*}
P(x_1,\cdots,x_n)=\sum_{i_1,\cdots,i_n=1}^{N}a_{i_1 \cdots i_n}x_1^{i_1}\cdots x_n^{i_n},
\end{eqnarray*}
where $a_{i_1 \cdots i_n}\in {\bf Z}$, $x_i \in {\bf R}$. For $P(x_1,\cdots,x_n)$, define its UD polynomial $P_{UD}:\Omega^n \to \Omega$ by
\begin{eqnarray*}
P_{UD}(X_1,\cdots,X_n)=\Max_{i_1,\cdots,i_n=1}^{N}\Bigl(\mbox{UD}(a_{i_1 \cdots i_n})+i_1 X_1+\cdots +i_n X_n \Bigr),
\end{eqnarray*}
where $X_i\in \Omega$.

More generally, let $F:{\bf R}^n \to {\bf R}$ a rational function given by
\begin{eqnarray*}
F(x_1,\cdots,x_n)=\frac{P(x_n,\cdots,x_1)}{Q(x_n,\cdots,x_1)},
\end{eqnarray*}
where $P$ and $Q$ are polynomial functions of finite degree.
Then, we define its UD equation $F_{UD}:\Omega^n \to \Omega$ by 
\begin{eqnarray}\label{eqn: UD equation}
F_{UD}(X_1,\cdots,X_n)=P_{UD}(X_1,\cdots,X_n)-Q_{UD}(X_1,\cdots,X_n).
\end{eqnarray}

\noindent{\it Remark.} 
(\ref{eqn: UD equation}) is just obtained by the following transformations:
\begin{eqnarray*}
&&x_i \times y_j \longrightarrow X_i + Y_j, \\
&&x_i / y_j \longrightarrow X_i - Y_j, \\
&&x_i + y_j \longrightarrow \mbox{Max}(X_i, Y_j), \\
&&x_i - y_j \longrightarrow \mbox{Max}(X_i, Y_j+ \eta ), \\
&&a_{i_1 \cdots i_n} \longrightarrow \mbox{UD}(a_{i_1 \cdots i_n}).
\end{eqnarray*}

\begin{theorem}\label{theorem: periodic equation}
Suppose a periodic recursive equation
\begin{eqnarray}\label{eqn; recursive equation}
x_{n+1}=F(x_1,\cdots,x_n)=\frac{P(x_n,\cdots,x_1)}{Q(x_n,\cdots,x_1)}
\end{eqnarray}
on ${\bf R}$ has period $k$ (i.e. $x_{n+k}=x_{n}$), then its UD periodic recursive equation 
\begin{eqnarray*}
X_{n+1}=F_{UD}(X_1,\cdots,X_n)=P_{UD}(X_1,\cdots,X_n)-Q_{UD}(X_1,\cdots,X_n)
\end{eqnarray*}
on $\Omega$ also has period $k$ (i.e. $X_{n+k}=X_{n}$).
\end{theorem}
\noindent{\it Proof.}
Substituting $x_n=e^{R_\epsilon(X_n)/\epsilon}$ into (\ref{eqn; recursive equation}), we have
\begin{eqnarray*}
e^{R_\epsilon(X_{n+1})/\epsilon}
&=&\frac{P(e^{R_\epsilon(X_n)/\epsilon},\cdots,e^{R_\epsilon(X_1)/\epsilon})}{Q(e^{R_\epsilon(X_n)/\epsilon},\cdots,e^{R_\epsilon(X_1)/\epsilon})}\\
&=&\frac{\sum_{i_1,\cdots,i_n=1}^{N}a_{i_1 \cdots i_n}e^{(i_1R_\epsilon(X_1)+\cdots +i_nR_\epsilon(X_n))/\epsilon}}{\sum_{i_1,\cdots,i_n=1}^{N}b_{i_1 \cdots i_n}e^{(i_1R_\epsilon(X_1)+\cdots +i_nR_\epsilon(X_n))/\epsilon}}\\
&=&\frac{\sum_{i_1,\cdots,i_n=1}^{N}e^{(\epsilon\log a_{i_1 \cdots i_n}+R_\epsilon(i_1X_1+\cdots i_nX_n))/\epsilon}}{\sum_{i_1,\cdots,i_n=1}^{N}e^{(\epsilon\log b_{i_1 \cdots i_n}+R_\epsilon(i_1X_1+\cdots i_nX_n))/\epsilon}}\\
&=&\frac{\sum_{i_1,\cdots,i_n=1}^{N}e^{R_\epsilon(U(a_{i_1 \cdots i_n})+i_1X_1+\cdots i_nX_n)/\epsilon}}{\sum_{i_1,\cdots,i_n=1}^{N}e^{R_\epsilon(U(b_{i_1 \cdots i_n})+i_1X_1+\cdots i_nX_n)/\epsilon}}.
\end{eqnarray*}
Here, in the last line, we have used 
\begin{eqnarray*}
\epsilon\log a_{i_1 \cdots i_n}&=&
\epsilon\log \underbrace{e^{0/\epsilon}+\cdots+e^{0/\epsilon}}_{a_{i_1 \cdots i_n}}\\
&=&R_\epsilon( (\underbrace{0,\cdots,0}_{a_{i_1 \cdots i_n}}))\\
&=&R_\epsilon(0+[\underbrace{0,\cdots,0}_{a_{i_1 \cdots i_n}-1}])\\
&=&R_\epsilon(U(a_{i_1 \cdots i_n})).
\end{eqnarray*}
Then, we have
\begin{eqnarray*}
&&R_\epsilon(X_{n+1})\\
&=&\Maxe_{i_1,\cdots,i_n=1}^{N}
(R_\epsilon(U(a_{i_1 \cdots i_n})+i_1X_1+\cdots i_nX_n))
-\Maxe_{i_1,\cdots,i_n=1}^{N}
(R_\epsilon(U(b_{i_1 \cdots i_n})+i_1X_1+\cdots i_nX_n)) \\
&=&R_\epsilon\Bigl(\Max_{i_1,\cdots,i_n=1}^{N}
(U(a_{i_1 \cdots i_n})+i_1X_1+\cdots i_nX_n)
-\Max_{i_1,\cdots,i_n=1}^{N}(U(b_{i_1 \cdots i_n})+i_1X_1+\cdots i_nX_n)\Bigr).
\end{eqnarray*}
Hence, we finally obtain
\begin{eqnarray*}
\mbox{R}_\epsilon(X_{n+1})=\mbox{R}_\epsilon(P_{UD}(X_1,\cdots,X_n)-Q_{UD}(X_1,\cdots,X_n)).
\end{eqnarray*}
On the other hand, we have $\mbox{R}_\epsilon(X_{n+k})=\mbox{R}_\epsilon(X_{n})$ from $x_{n+k}=x_{n}$. This implies $X_{n+k}=X_{n}$.
$\square$
\newline

Next, we will give some examples.
\begin{example}
The periodic recursive equation
\begin{eqnarray*}
x_{n+1}=\frac{x_n}{x_{n-1}}
\end{eqnarray*}
has period 5. Then, from Theorem \ref{theorem: periodic equation}, we find that its UD equation 
\begin{eqnarray}\label{eqn:example1}
X_{n+1}=\mbox{Max}(X_n, 0)-X_{n-1}
\end{eqnarray}
 also has period 5. For example, we have
\begin{eqnarray}
&&X_1=1,~~~X_2=5,~~~X_3=4+[5],~~~X_4=-1+[4,5],\nonumber\\
&&X_5=-4+[1,5,6]-[5],~~~X_6=1,~~~X_7=5.\label{eqn:example2}
\end{eqnarray}
\end{example}
\noindent{\it Remark.} If $X_i \in \Omega_0$ for all $i$, then we can act the projection map $\mbox{P}:\Omega_0 \to \bf{Z}$ for both sides of equation (\ref{eqn:example1}). Then, we have 
\begin{eqnarray*}
Y_{n+1}=\mbox{max}(Y_n, 0)-Y_{n-1},
\end{eqnarray*}
where $Y_i=P(X_i)$. This equation also has period 5. For example, 
\begin{eqnarray*}
&&Y_1=1,~~~Y_2=5,~~~Y_3=4,~~~Y_4=-1,\\
&&Y_5=-4,~~~Y_6=1,~~~Y_7=5,
\end{eqnarray*}
which correspond to (\ref{eqn:example2}).

\begin{example}
The periodic recursive equation
\begin{eqnarray*}
x_{n+1}=\frac{x_n-1}{x_{n}+1}
\end{eqnarray*}
has period 4. Then, from Theorem \ref{theorem: periodic equation}, we find that its UD equation   
\begin{eqnarray}\label{eqn:example3}
X_{n+1}=\mbox{Max}(X_n, \eta)-\mbox{Max}(X_n, 0)
\end{eqnarray}
also has period 4. For example, 
\begin{eqnarray}
&&X_1=2,~~~X_2=[2+\eta]-[2],~~~X_3=-2+\eta,\nonumber\\
&&X_4=\eta+[2]-[2+\eta],~~~X_5=2.\label{eqn:example4}
\end{eqnarray}
\end{example}
\noindent{\it Remark.} We cannot act the projection map $\mbox{P}:\Omega_0 \to \bf{Z}$ for both sides of equation (\ref{eqn:example3}) and (\ref{eqn:example4}), since all the variables $X_i$ do not belong to $\Omega_0$. Therefore, this is a new type of UD equations, which cannot be obtained by the usual ultra-discretization procedure.


\begin{thebibliography}{99}
  \bibitem{Hirota} R. Hirota, K. Kimura and H. Yahagi: How to find the conserved quantities of nonlinear discrete equations, J. Phys. A {\bf 34} (2001) 10377-10386.
  \bibitem{Hirota T} R. Hirota, S. Tsujimoto: Conserved Quantities of a Class of Nonlinear Difference-Difference Equations, J. Phys. Soc. Jpn. {\bf 64} (1995) 3125-3127. 
  \bibitem{TakahashiSatsuma} D. Takahashi and J. Satsuma: A Soliton Cellular Automaton, J. Phys.Soc. Jpn. {\bf 59} (1990) 3514-3519.
  \bibitem{Takahashi2} D. Takahashi, T. Tokihiro, B.Grammaticos, Y. Ohta, A Ramani: Constructing Solutions to the Ultra-Discrete Painleve Equations, J. Phys. A {\bf 30} (1997) 7953-7966. 
  \bibitem{Tokihiro2} T. Tokihiro, A. Nagai and J. Satsuma: Proof of solitonical nature of box and ball system by Means of Inverse Ultra-Discretization, Inverse Problems {\bf 15} (1999) 1639-1662.
  \bibitem{Tokihiro T M }T. Tokihiro, D. Takahashi, J. Matsukidaira: Box and Ball system as a realization of ultradiscrete nonautonomous KP equation, J. Phys. A{\bf 33} (2000) 607-619.  
  \bibitem{Tokihiro}T. Tokihiro, D. Takahashi, J. Matsukidaira and J. Satsuma: From Soliton Equations to Integrable Cellular Automata through a Limiting Procedure, Phys. Rev. Lett. {\bf 76} (1996) 3247-3250.
  \bibitem{Torii T S} M. Torii, D. Takahashi and J. Satsuma: Combinatorial Representation of Invariants of Soliton Cellular Automaton, Physica D {\bf 92} (1996) 209-220. 
   \bibitem{Hatayama H I K T T} G. Hatayama, K. Hikami, R. Inoue, A. Kuniba, T. Takagi and T. Tokihiro: The $A_M^{(1)}$ automata related to crystals of symmetric tensors, J. Math. Phys. {\bf 42} (2001) 274-308.
  \bibitem{Hatayama K T} G. Hatayama, A. Kuniba and T. Takagi: Simple Algorithm for Factorized Dynamics of $g_n$-Automaton, J. Phys. A {\bf 34} (2001) 10697-10705.
  
  
   
  
\end{thebibliography}
\end{document}